\documentclass[
  ,final            % use final for the camera ready runs
  ]
  {aipproc}

\layoutstyle{8x11single}

\usepackage{graphicx}

\newcommand{\nope}[1]{}
\newcommand{\ppbar}{$p \bar p$\ }
\newcommand{\met}{$\slash\hspace{-6pt}E_T$}
\newcommand{\xsecss}[2]{$\sigma_{t\bar t}=(#1 #2(\mathrm{stat.+syst.}))$\ pb\ }
\newcommand{\xsec}[3]{$\sigma_{t\bar t}=(#1 #2(\mathrm{stat.}) #3(\mathrm{syst.}))$\ pb\ }
\newcommand{\xsecl}[4]{$\sigma_{t\bar t}=(#1 #2(\mathrm{stat.}) #3(\mathrm{syst.})\pm#4(\mathrm{lumin.}))$\ pb\ }
\newcommand{\etal}{{\it et al.}}

\begin{document}

\title{Top Quark Production and Decay Properties at the Tevatron}


\author{M.Weber for the CDF and D0 collaborations} {
address={Fermilab, P.O. Box 500, Batavia, IL 60510, USA}
}

\begin{abstract}
The latest results from the CDF and D0 collaborations on the top-quark pair-production cross section and limits on electroweak production are presented.
Included are measurements of properties of the top quark such as charge, lifetime, and the decay branching ratio $t\rightarrow Wb$.
In addition to measurements about the top quark, the selected event samples are used to study the helicity of the $W$ boson and to search for additional exotic quarks ($t'$) and resonances in the $t \bar t$ invariant mass spectrum.

\end{abstract}

\maketitle

%%%%%%%%%%%%%%%%%%%%%%%%%%%%%%%%%%%%%%%%%%%%
%% MAINMATTER
%%%%%%%%%%%%%%%%%%%%%%%%%%%%%%%%%%%%%%%%%%%%

\section{Introduction}
The top quark was discovered in 1995 by the CDF and D0 collaborations at the Fermilab Tevatron Collider \cite{topdiscovery}.
The CDF and D0 collaborations are currently taking data in Run 2 of the Tevatron.
The increased luminosity and higher collision energy of $\sqrt{s}$ = 1.96 TeV allows for precise measurement of top quark production and decay properties.
The CDF and D0 detectors are described in Ref. \cite{cdf} and \cite{d0}.
The top quark is by far the heaviest particle found to date ($m_t=(171.4\pm2.1)$ GeV \cite{massavg}) and is also the quark with the smallest uncertainty on its mass.
Due to its high mass, it plays a central role in the standard model (SM).
This article focusses on the production cross sections, decay, and properties of the top quark.
The mass measurements are discussed in a separate article in these proceedings \cite{unki}.

\section{Top quark production}

The top quark can be pair produced in \ppbar collisions via the strong interaction and singly via the electroweak interaction in the SM.
The $t \bar t$ cross section is eleven orders of magnitude lower than the inelastic \ppbar cross section and several orders of magnitude lower than $b$ quark and$ $W and $Z$ boson production, which poses a significant challenge to top quark identification.
\nope{
as can be seen in Fig.\ \ref{xs}.
\begin{figure}
  \includegraphics[height=.3\textheight]{xs}
  \label{xs}
  \caption{Cross sections}
\end{figure}	
}
In the SM, the top quark decays almost exclusively into a $W$ boson and a $b$ quark, where the $W$ boson decays either hadronically or leptonically.
Identification of top quarks require the identification of jets, in particular jets from $b$ quarks, muons, electrons, and neutrinos.
The identification is based on the fact that decay products have hight transverse momenta and good angular separation in the lab frame.
The principal algorithm used to identify $b$ quark jets looks for the presence of charged tracks significantly displaced from the primary vertex coming from the decay of $B$ or $D$ mesons, which have finite lifetime (lifetime tagging).

\subsection{Top quark production via the strong interaction}

Top quarks can be produced in pairs via the strong interaction.
At the Tevatron collision energy of $\sqrt{s}=1.96$ TeV the quark-annihilation diagrams make up 85\% of the cross section, while gluon fusion accounts for 15\%.
Theoretical calculations predict a $t\bar t$ production cross section of $6.7^{+0.9}_{-0.7}$ pb at a top quark mass of 175 GeV \cite{kidonakis,cacciari}.
The top quark candidate events are classified according to the $W$ boson decay mode.
Each top quark decays into a $W$ boson, which can decay either hadronically or leptonically.
Decay channels include di-lepton (both $W$ bosons decay leptonically), semi-leptonic (mixed), and all-jets (only hadronic decays). 
The $t \bar t$ production cross section has been measured by the CDF and D0 collaborations in all decay channels.
Selected results are presented here.

\begin{figure}
  \label{xsecplot}
  \includegraphics[height=.3\textheight]{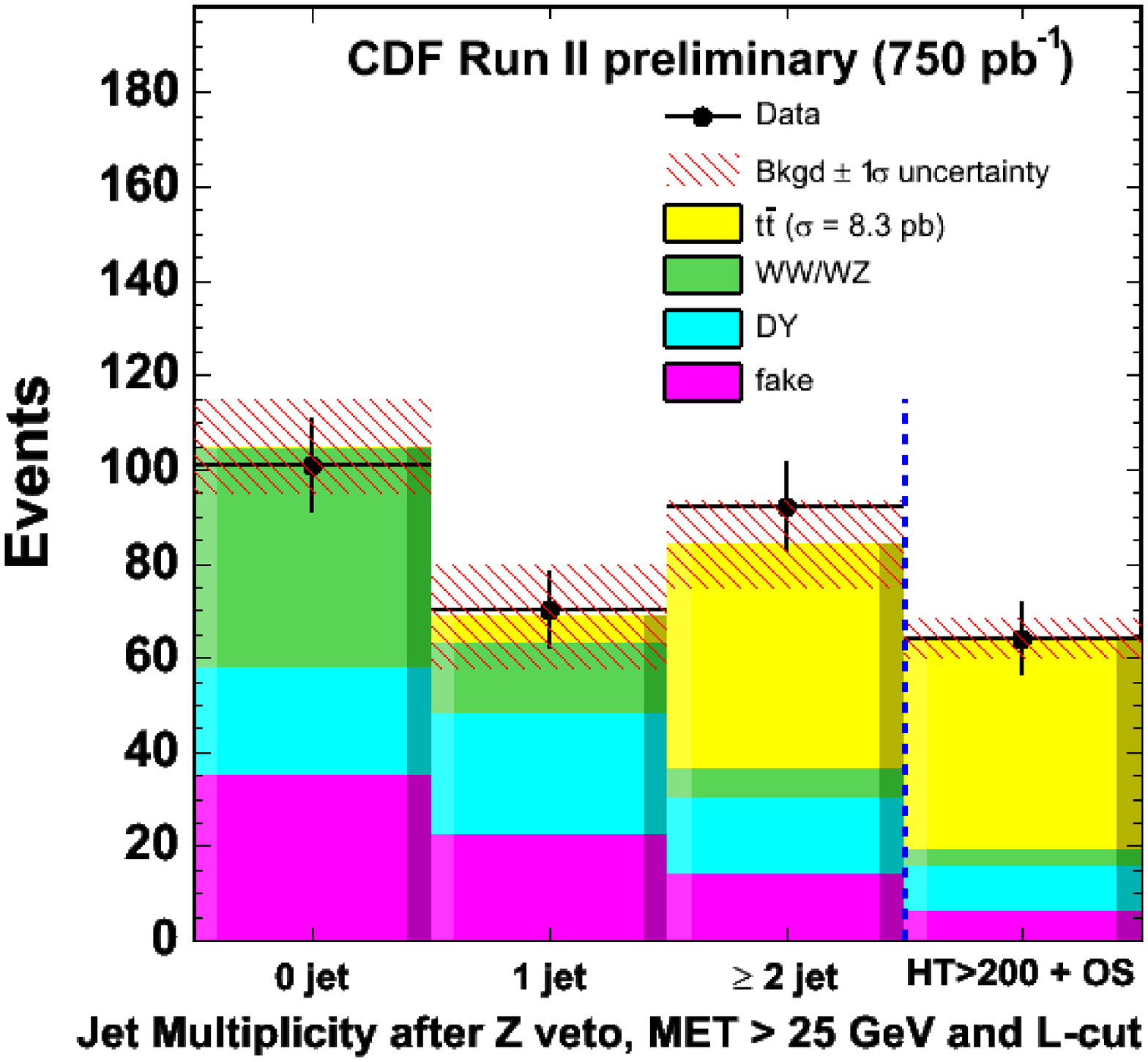}
  \hspace*{10mm}
  \includegraphics[height=.3\textheight]{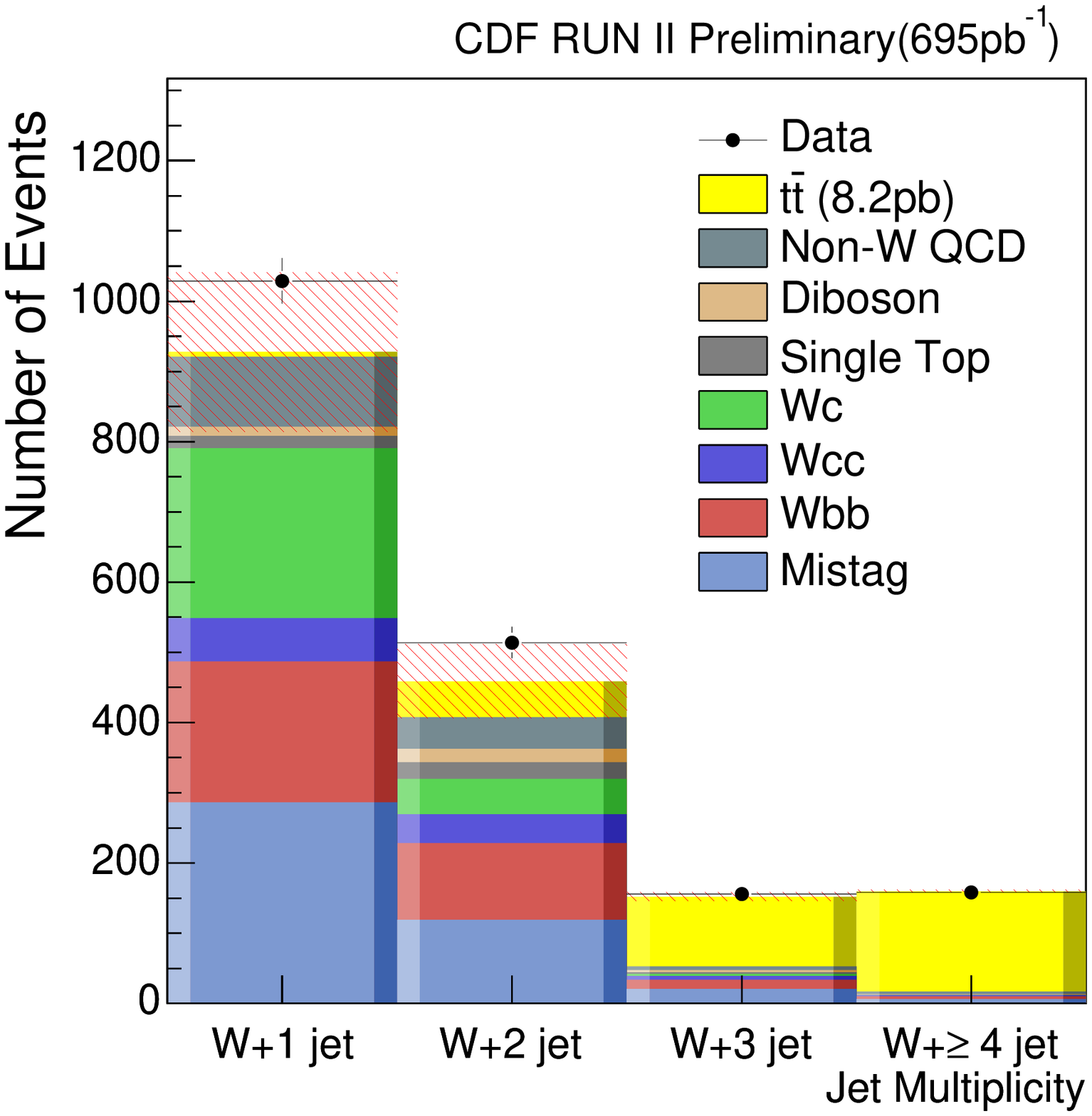}
  \caption{
    {\bf Left:}
    Candidate events (black points) from the CDF di-lepton sample by jet multiplicity.
    The histograms represent the background expectation for 750 pb$^{-1}$ of integrated luminosity.
    The top light histogram is the expected contribution from signal with $\sigma_{t\bar t}=8.3$ pb as measured in this analysis.
    The shaded area is the uncertainty in the total background estimate.
    {\bf Right:}
    Number of events in the CDF semi-leptonic event selection as a function of jet multiplicity. 
    For events with at least three jets an additional requirement is made of $H_T>200$ GeV.)
    The histograms represent the background expectation.
    The black points are the observation in data.
    The light-colored histogram is the contribution from $t\bar t$ normalized to the measured cross section.
  }
\end{figure}

The di-lepton channel is characterized by the presence of two isolated high $p_T$ leptons, two high $p_T$ $b$-jets, and a large missing transverse energy (\met) from the neutrinos.
The  background contributions due to instrumental effects are multi-jet production, $W$ boson with additional jets ($W$+jets), and $Z\rightarrow$ll events with mis-measured \met, misidentified jets or misidentified leptons.
These backgrounds are estimated from data.
Physics background include Z$\rightarrow\tau\tau$ where the $\tau$ leptons decay leptonically and $WW$/$WZ$ (di-boson) production and are estimated from monte carlo (MC) simulation.
The di-lepton channel has the advantage of clean lepton identification and therefore high signal to background rations.
However, the branching ratio of 4\% (excluding $\tau$ lepton decay modes) is low.
An analysis by CDF includes events with all combinations of electron and muon decays of the $W$ bosons.
Figure \ref{xsecplot} shows the number of candidate events in data compared to the expectation from background and signal $t \bar t$.
The final selection requires the leptons to have opposite sign charge and a cut on $H_T>200$ GeV.
A preliminary cross section of \xsecl{8.3}{\pm1.5}{\pm1.0}{0.5} is extracted.

To increase the efficiency of the identification the D0 collaboration has selected a sample where only one lepton is required to be fully reconstructed allowing for only a track identified from the decay of the other charge $W$ boson.
Events with one $W$ boson decaying into an electron and the other into a muon (electron-muon events) are rejected.
The signal purity lost by loosening the lepton identification is recovered by the requirement of at least one $b$-jet to be identified (tagged) by a lifetime tagger.
The measured cross section is then combined with the cross section from electron-muon events and yields a preliminary \xsecl{8.6}{^{+1.9}_{-1.7}}{\pm1.1}{0.6}.
The statistical error in this result is reduced by 15\% compared to the D0 result with full lepton identification, keeping the systematic error comparable.

In the semi-leptonic decay channel the identification of the lepton still provides a reasonably high signal to background ratio, although not as clean as the di-lepton channel.
The branching ratio is 29\% (not including the $\tau$+jets branching ratio) and therefore significantly higher than the di-lepton channel.
This makes the semi-leptonic channel the golden channel for top-quark property measurements.
The dominant background processes are $W$+jets and multi-jet production.

CDF measures the cross section in the semi-leptonic channel by selecting a sample of events with an isolated electron (muon) with transverse energy (transverse momentum) $E_T(p_T)>20$ GeV, \met>20 GeV and jets with $E_T>15$ GeV and $|\eta|<2$.
At least one of the jets is also required to be tagged as a $b$-jet by a lifetime tagger.
Further separation between signal $t \bar t$ and $W$+jets background is obtained by employing a cut on $H_T$.
Efficiency and physics backgrounds are estimated from MC simulation.
Instrumental backgrounds are estimated from data.
The selected number of events from 695 pb$^{-1}$ of data are shown in Fig.\ \ref{xsecplot} as a function of the jet multiplicity in the event.
A preliminary $t \bar t$ cross section is extracted from events with at least three jets and is measured to be \xsec{8.2}{\pm0.6}{\pm1.0} for a top quark mass of 175 GeV.
For the $t \bar t$ cross section measurement in the semi-leptonic channel D0 splits the sample by lepton flavor, jet multiplicity, and number of lifetime tagged jets.
The combined preliminary measurement yields \xsecss{8.1}{^{+1.3}_{-1.2}} from collisions with 360 pb$^{-1}$ of integrated luminosity.

\begin{figure}
  \includegraphics[height=.3\textheight]{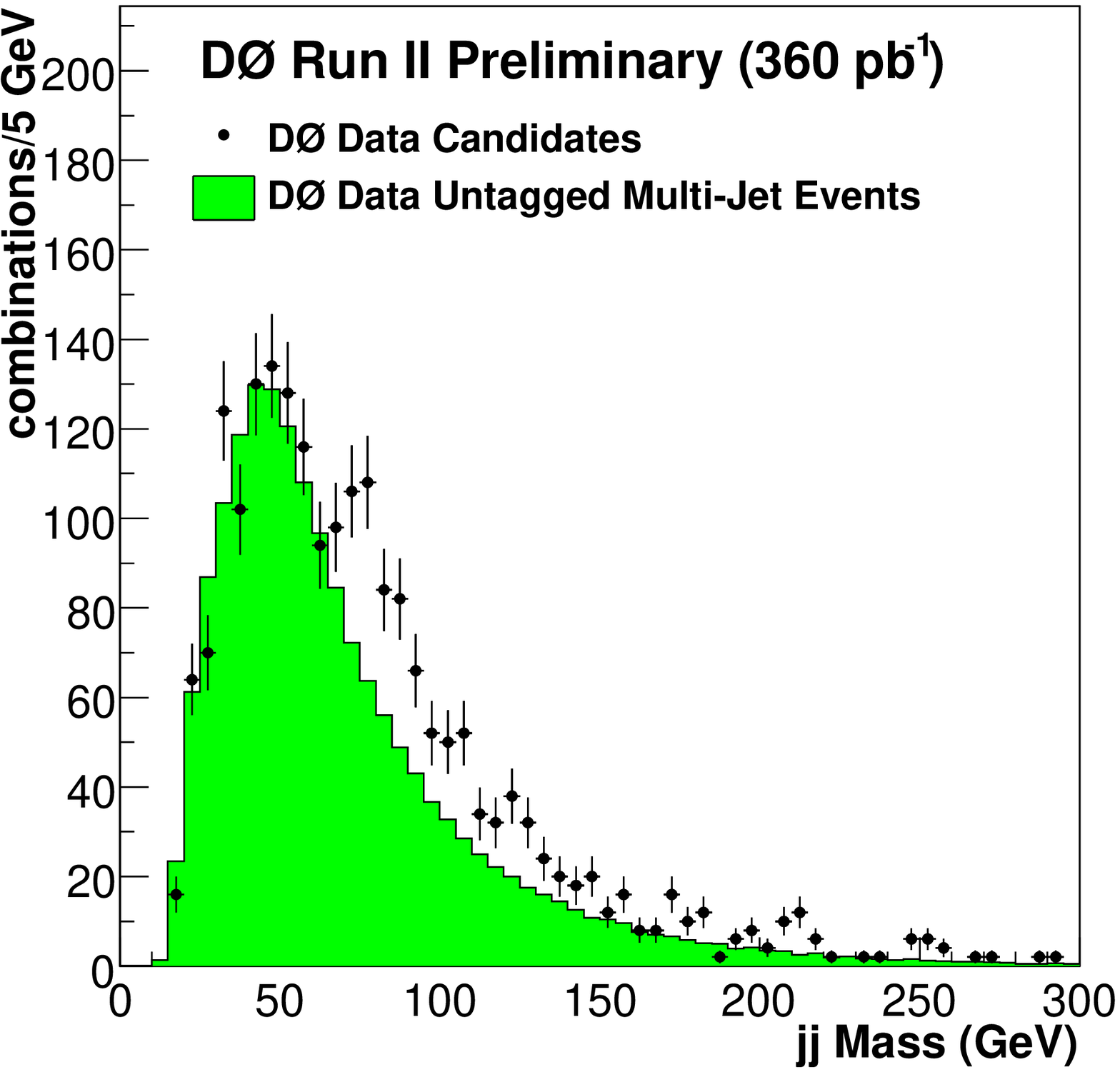}
  \hspace*{10mm}
  \includegraphics[height=.3\textheight]{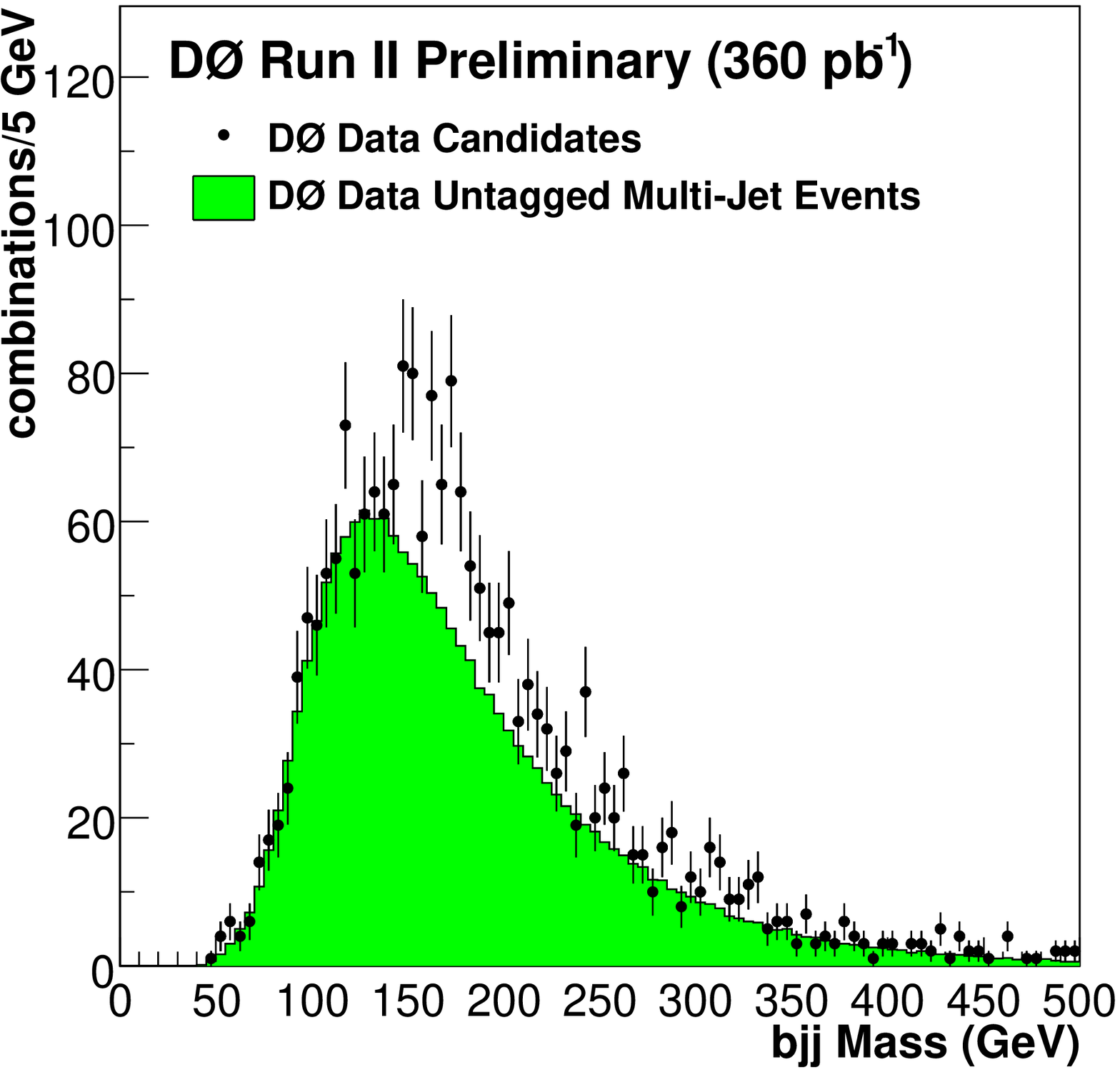}
  \label{alljet}
  \caption{{\bf Left:} The di-jet mass spectrum for all pairs of non $b$-tagged jets from the D0 all-jets cross section measurement.
    {\bf Right:} The three-jet mass spectrum for all combinations
of one $b$-tagged jet and two non $b$-tagged jets.
The green-shaded histogram overlaid on the data points is the distribution
from the background sample.}
\end{figure}

The $t \bar t$ cross section is also measured in the events where both $W$ boson decay hadronically.
This final state is characterized by six high $p_T$ jets, two of which are $b$-jets.
The dominant background in this channel is multi-jet production.
Although the branching ratio of 46\% into this channel is relatively large the background processes have much larger cross sections making the signal hard to identify.
D0 extracts a signal by building invariant masses from two-jet and three-jet combinations in each event.
The jets for the di-jet invariant mass are required not to be tagged by a lifetime tagger (light jets).
For the three-jet invariant mass one of the jets has to be identified as a $b$-jet.
Figure \ref{alljet} shows the di-jet (left) and three-jet (right) mass distributions.
Overlaid are the background distributions also obtained from data by assigning a $b$-flavor to a jet at random.
The excess of events in the di-jet spectrum reveals the $W$ boson, while in the three-jet distribution the excess coming from top quarks can be seen.
The preliminary cross section measured using 360 pb$^{-1}$ of data is \xsec{12.1}{\pm4.9}{\pm4.6}.

\begin{figure}
  \includegraphics[height=.3\textheight]{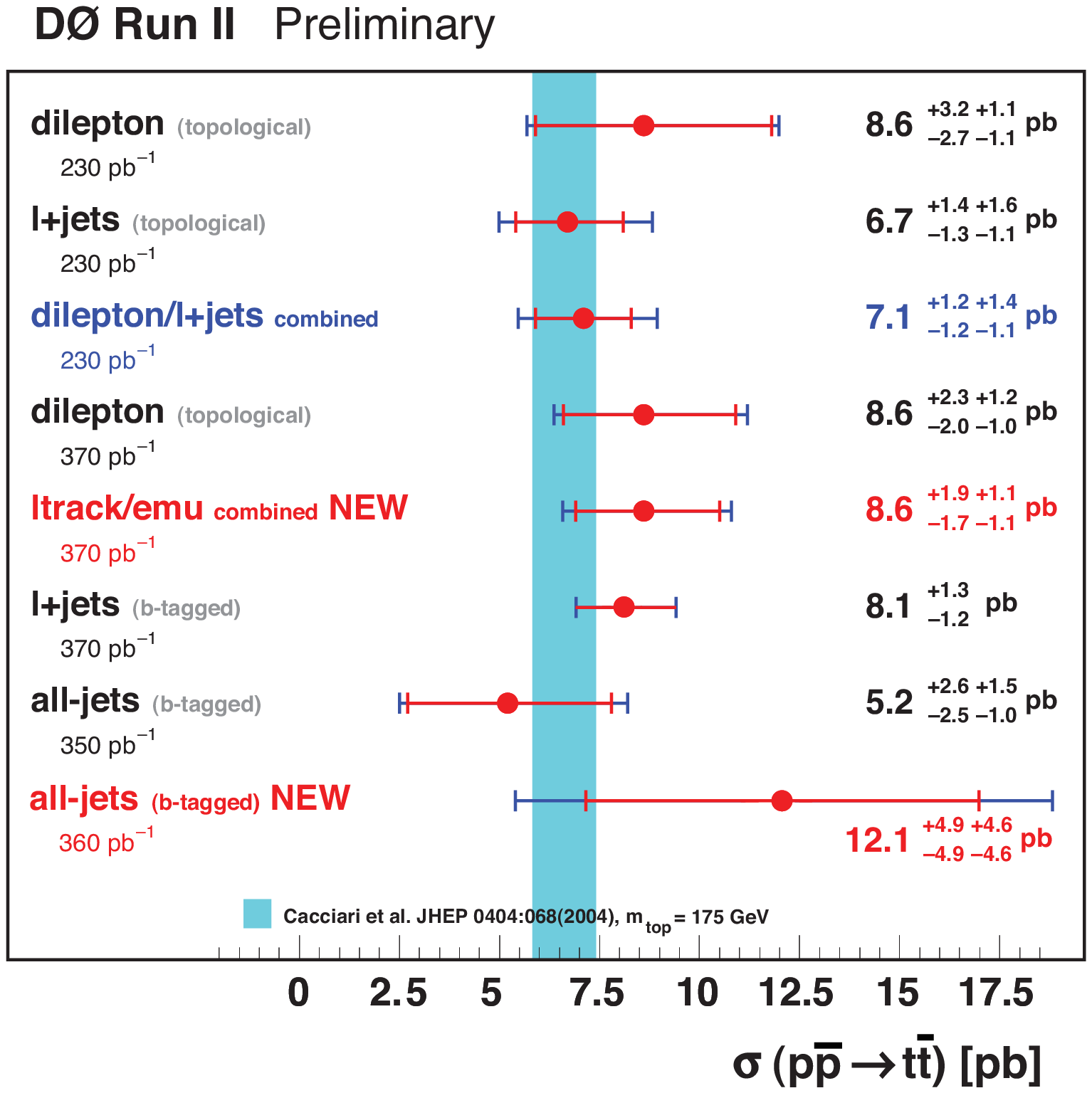}
  \hspace*{12mm}
  \includegraphics[height=.3\textheight]{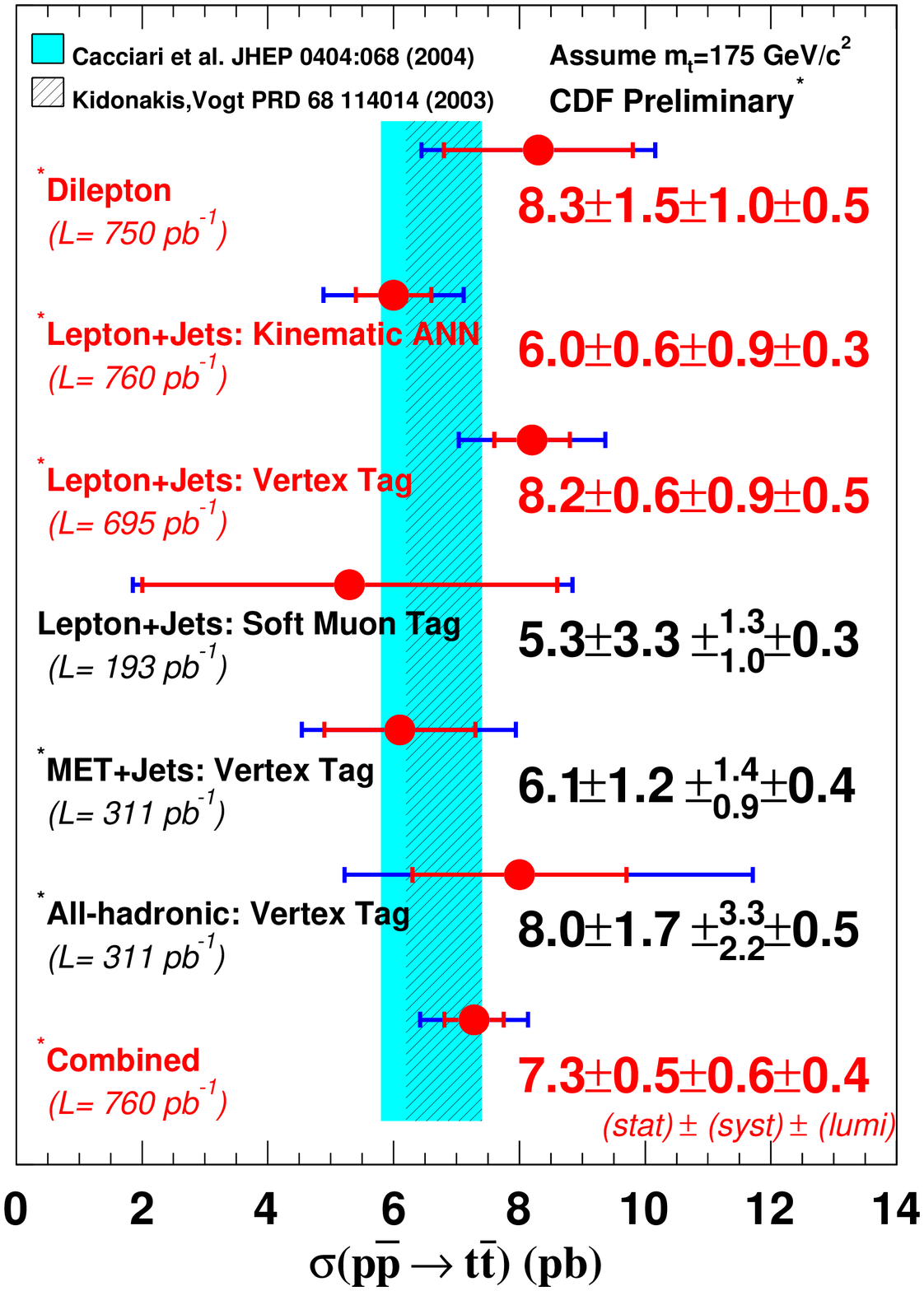}
  \label{topxs}
  \caption{Compilation of $t \bar t$ production cross section measurements by D0 (left) and CDF (right).}
\end{figure}

A compilation of cross section measurements by CDF and D0 is shown in Fig.\ \ref{topxs}.
  
\subsection{Top quark production via the electroweak interaction}
While top quark pair production via the strong interaction is well established and measured, top quarks may be produced individually via the electroweak interaction.
No evidence for this production channel has not been found to date.
Measurement of single top quark production directly access the CKM matrix element $|V_{tb}|$ and properties of the $tWb$ vertex.
At the Tevatron the SM predicts single top quark production to occur from two dominant sets of diagrams, the $s$ channel with an expected cross section of $1.98\pm0.25$ pb and the $t$ channel diagrams with $0.88\pm0.11$ pb \cite{harris}. 
Associated production of top quarks is expected to have negligible cross section.
 Single top quark production is searched for in events with the top quark decaying into a $W$ boson and a $b$ quark with the $W$ boson decaying leptonically.
A lifetime tagger is used to identify the $b$ quark jet.
The major challenge of the analyses is separating the single top signal from the $t \bar t$ and $W$+jets background.
Multi-variate statistical methods including multi-variate discriminants, artificial neural networks (ANN), and decision trees are used by both D0\cite{poster} and CDF.
Separate trainings on MC simulated events are performed for the separation of single top signal to $t \bar t$ and $W$+jets.
\nope{
  An example of the separations from the D0 discriminant analysis can be found in Fig.\ \ref{d0st}.
\begin{figure}
  \includegraphics[height=.25\textheight]{d0sta}
  \includegraphics[height=.25\textheight]{d0stb}
  \label{d0st}
  \caption{Output of the D0 multi-variate likelihood discriminants for separation of single top signal to $t \bar t$ (left) and $W$+jets (right) background.
  The separate contribution from simulation is shown as histograms and compared to the data point.
  The expected single top signal is multiplied by a factor of 10 and shown as lines.
}
\end{figure}	
}

\begin{figure}
  \includegraphics[height=.3\textheight]{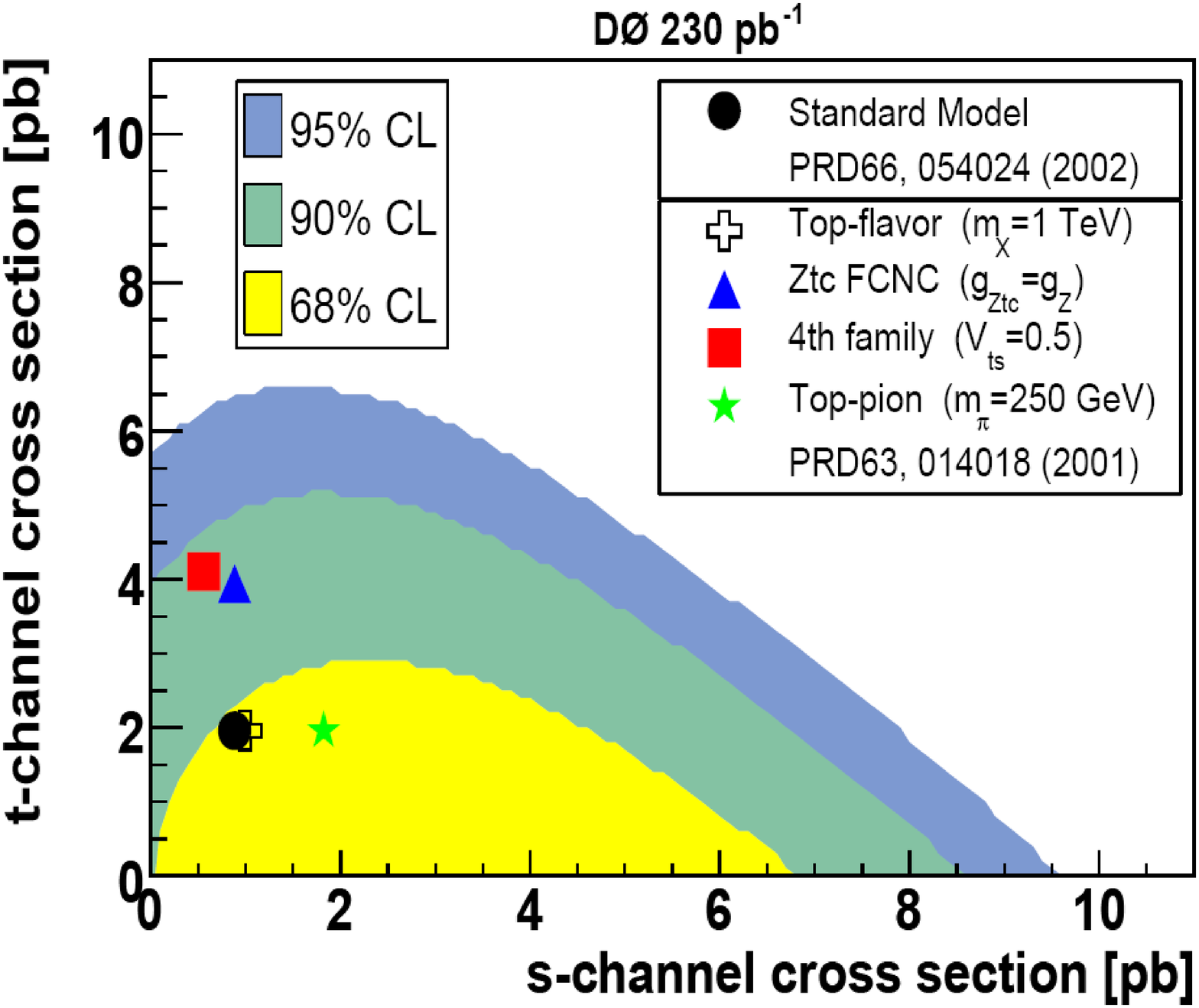}
  \label{stnp}
  \caption{
    Allowed regions for single top quark production in the ($s$-channel, $t$-channel) plane for the D0 neural network analysis.
    Several model predictions are also shown in the plot.
  }
\end{figure}

A likelihood fit is made to the discriminant output as a function of input single top cross section.
Signal significance or limits are extracted from the posterior probability density distribution.
No evidence for single top quark production is found by CDF or D0.
With 370 pb$^{-1}$ of integrated luminosity the limits from D0 (likelihood discriminant) at 95\% C.L. are $\sigma_t<5.0$ pb for $t$-channel and  $\sigma_t<4.4$ pb for $s$-channel.
Earlier limits by D0 are published in Ref. \cite{d0st}.
The limits with 695 pb$^{-1}$ from CDF (neural network analysis) at 95\% C.L. are $\sigma_t<3.2$ pb for $s$-channel, $\sigma_t<3.1$ pb for $t$-channel and $\sigma_t<3.4$ pb for a combined search.
Earlier limits by CDF are published in Ref. \cite{cdfst}.
Although single top quark production has yet to be observed, the current limits test models of physics beyond the SM \cite{tait}.
Figure \ref{stnp} shows the allowed region for single top quark production in the 2D $s$-channel - $t$-channel plane for the D0 neural network analysis.
Several model predictions for physics beyond the SM are also shown.
Both CDF and D0 continue to refine their analysis techniques to ensure the necessary sensitivity to find evidence for single top quark production with the data available in the very near future.

\section{Top quark decay and properties}

\subsection{$B(t\rightarrow Wb)/B(t\rightarrow Wq)$}
\begin{figure}
  \includegraphics[height=.25\textheight]{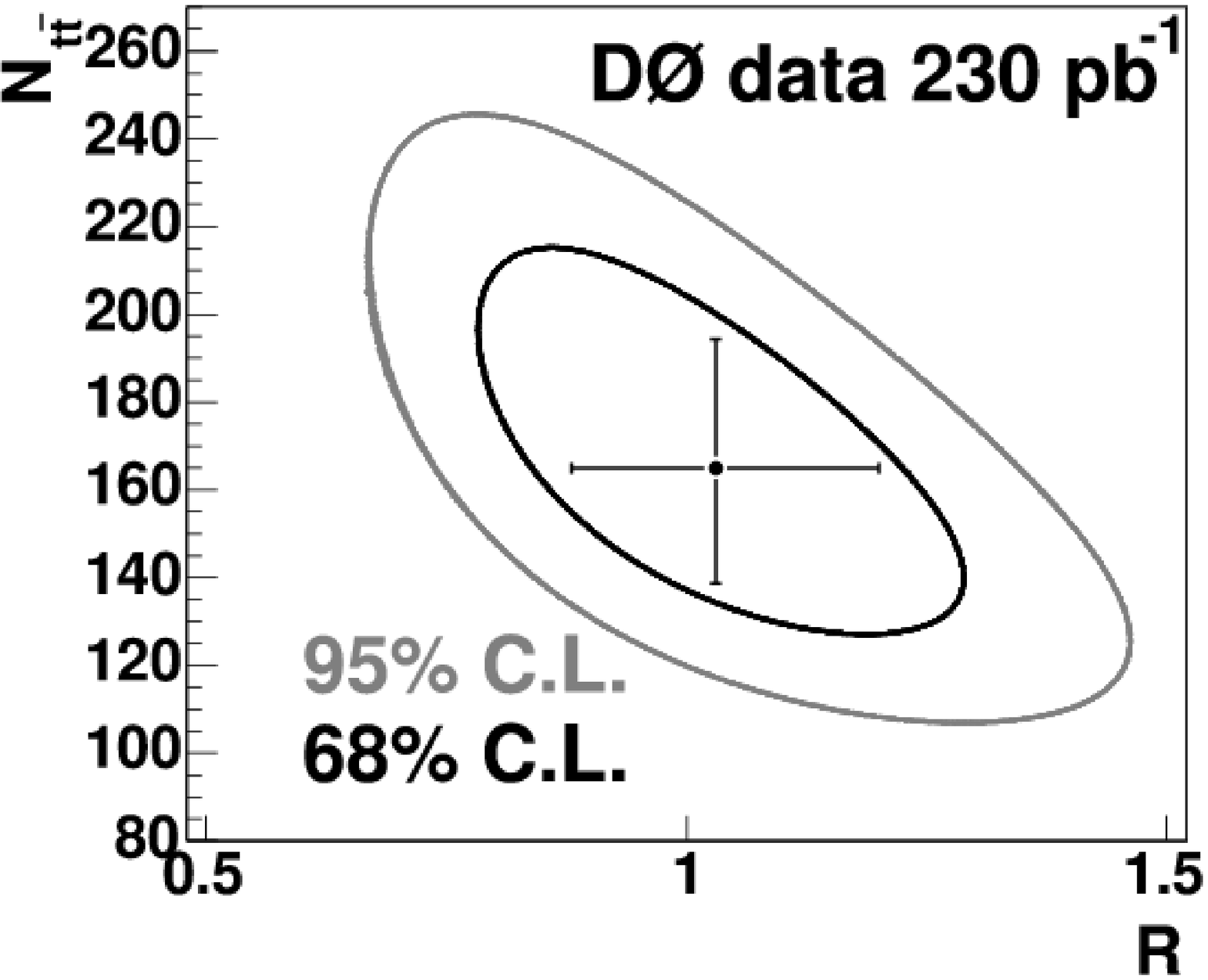}
  \hspace*{10mm}
  \includegraphics[height=.25\textheight]{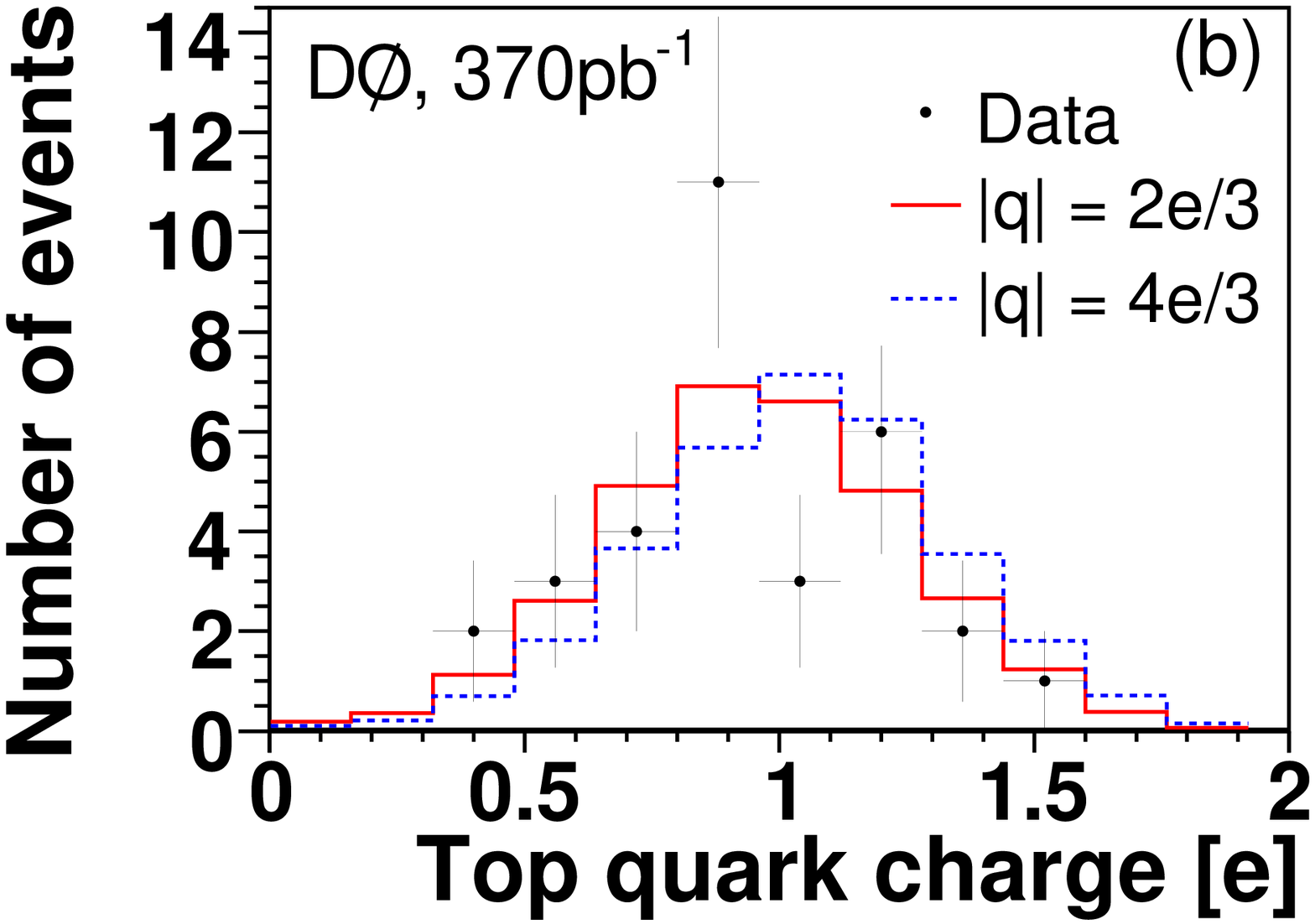}
  \label{bratiocharge}
  \caption{
    {\bf Left:}
    The 68\% and 95\% statistical confidence contours in the ($R$, $N_{t \bar t}$) plane.
  The point indicates the best fit to the data.
    {\bf Right:}
    The measured value of the top quark charge in 32 events selected by D0 compared to the expected distributions in the SM and exotic case.
  }  
\end{figure}

CDF and D0 measure the ratio $R=B(t\rightarrow Wb)/B(t\rightarrow Wq)$.
In the SM this ratio is tightly constrained assuming unitarity of the CKM matrix and exactly three quark generations to the interval 0.9980-0.9984 at 90\% CL.
Through the measurement of $R$, the SM and assumptions can be tested.
In addition, the cross section measurements using lifetime tagging to identify $b$-jets also assume $R$ to be unity.
A separate determination on $R$ can provide a model independent measurement of the cross section.
The measurement is done on a sample similar to the one used for the cross section measurement in the semi-leptonic channel.
$R$ is measured from a fit to the relative number of events with zero, one, and two $b$-tagged jets.
Tagging probabilities also affect these relative numbers and is determined from data control samples.
The cross section information comes from the sum excess of events over the background expectation.
The preliminary result from the simultaneous determination of cross section and $R$ on data from 230 pb$^{-1}$ of integrated luminosity is shown in Fig.\ \ref{bratiocharge}.
  $R$ is found to be $1.03^{+0.19}_{-0.17}$(stat.+syst.) and the cross section is \xsecss{7.9}{^{+1.7}_{-1.5}}.
Using a Bayesian method the lower limit on $R$ is found to be 0.64 at 95\% C.L.
CDF also measures $R$ and finds a value of $R=1.12^{+0.21}_{-0.19}(\mathrm{stat.})^{+0.17}_{-0.13}(\mathrm{syst.})$ \cite{cdfr}.

\subsection{$W$ boson helicity}
Given the $V-A$ structure of the electroweak coupling in the SM and the measured top quark mass, the fraction of $W$ bosons from top quark decay with left-handed, longitudinal, and right-handed polarizations are predicted to be $f_0=0.70$ and $f_{-}=0.30$ with an uncertainty of order 1\% and $f_+=\mathcal O(10^{-4})$ .
A measurement that departs from these values  would be a sign of physics beyond the SM.
For example, a $V+A$ term in the $tWb$ coupling would increase $f_+$ but leave $f_0$ unchanged.
Top quark decays with the $W$ boson decaying into an electron or a muon are used to measure the $W$ boson helicity.
To measure the helicity of the $W$ boson one looks at the distribution of angles between the charged lepton and the top quark (opposite to the $b$ quark) in the $W$ boson rest frame.
This helicity angle can be either directly measured or by using the approximation $\cos\theta^*\approx(2m^2_{\mathrm{lb}})/(m^2_t-m^2_W)-1$.
D0 uses the former approach and obtains $f_+=0.08\pm0.08$(stat.)$\pm0.06$(syst.) by combining events from semi-leptonic and di-lepton decays.
A third approach is to use the fact that the transverse momentum of the lepton in the laboratory frame depends on the $W$ boson helicity since left-handed $W$ bosons tend to emit the lepton opposite of their direction. 
CDF combines the result from this approach with the measurement from the $\cos\theta^*$ approximation and finds $f_0=0.74^{+0.22}_{-0.34}$(stat.+syst.) or $f_+<0.27$ at 95\% C.L.

\subsection{Top quark electric charge}
The electric charge is a fundamental quantity characterizing a particle.
In the SM the top quark is defined as having charge +2/3$e$.
In a possible extension to the SM an additional quark doublet $(Q_1,Q_4)_R$ with charges (-1/3$e$,-4/3$e$) is proposed \cite{chang}.
The true top quark in such models is too heavy to be observed at the Tevatron and the discovered top quark is indeed $Q_4$.
We can distinguish between the two model by measuring the top quark charge.
The measurement is done on a semi-leptonic sample of $t \bar t$ candidate events.
The charge of the top quark is measured from the charge of the lepton and the charge of the associated $b$-quark.
The lepton charge is measured from the curvature of the charged track in the magnetic field of the detector.
The $b$-quark charge is measured from a $p_T$ weighted sum of the charge tracks within jet cone.
One can distinguish between the $b$ and $\bar b$ on a statistical basis.
Figure \ref{bratiocharge} shows the reconstructed charges compared to the expectation from SM and -4/3$e$ scenario.
From a likelihood ratio fit the data is found to favor the SM and a preliminary exclusion of the -4/3$e$ model at 92\% C.L can be set.

\subsection{Top quark lifetime}
The top quark lifetime is constrained in the SM to be less than $10^{-24}$ s.
However, there is ample experimental room for long-lived top quark in the experimental data.
The top quark lifetime is measured in a sample of $t \bar t$ candidates selected by identifying an electron/muon, at least three jets, and large missing transverse energy.
One of the jets is required to be identified as a $b$ jet from a secondary vertex tag.
In such events, the impact parameter ($d_0$) of the lepton to the interaction vertex is typically expected to be very small.
The vertex location is constrained by the jets.
Large impact parameters would arise from longer-lived top quarks leading to displaced $W$ boson decay vertexes.
The distributions of lepton impact parameter $d_0$, only reflecting the resolution of the measurement is predicted using high momentum electron and muon tracks produced in $Z \rightarrow e^+e^-/\mu^+\mu^-$ events.
Distribution from non $t \bar t$ backgrounds are evaluated from the MC simulation.
The observed distribution in data is compared to the predicted distribution.
The preliminary result from 318 pb$^{-1}$ of CDF data shows $d_0$ distributions consistent with expectations and a limit on the top quark lifetime of $c\tau_t < 52.5$\ $\mu$m at 95\% C.L.
The analysis is also sensitive to a new long-lived background to $t \bar t$ or anomalous top quark production by a long-lived parent particle.

\section{Searches for new particles in top quark production and decay}
Many models beyond the SM predict new particles that couple preferentially to the top quark.
Two searches are presented here, a search for a narrow heavy resonance $X \rightarrow t \bar t$ and the search for an additional quark $t'$ with the same final states as the SM top quark.

\subsection{Search for a narrow resonance in $t\bar t$ production}
\begin{figure}
  \includegraphics[height=.3\textheight]{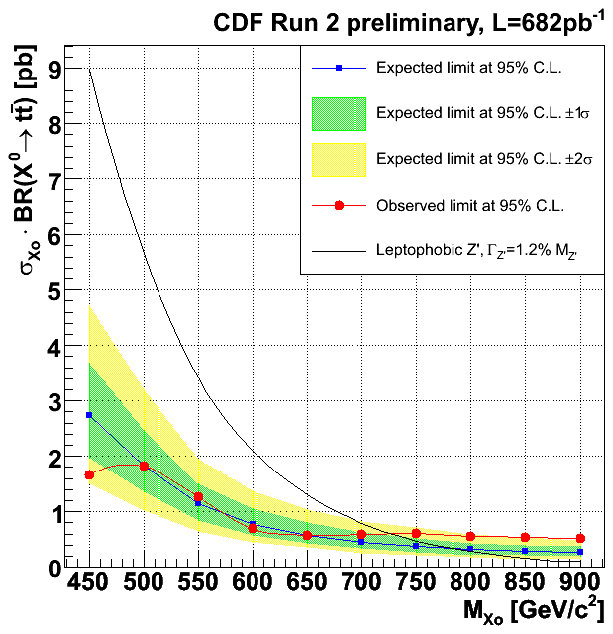}
  \hspace*{10mm}
  \includegraphics[height=.25\textheight]{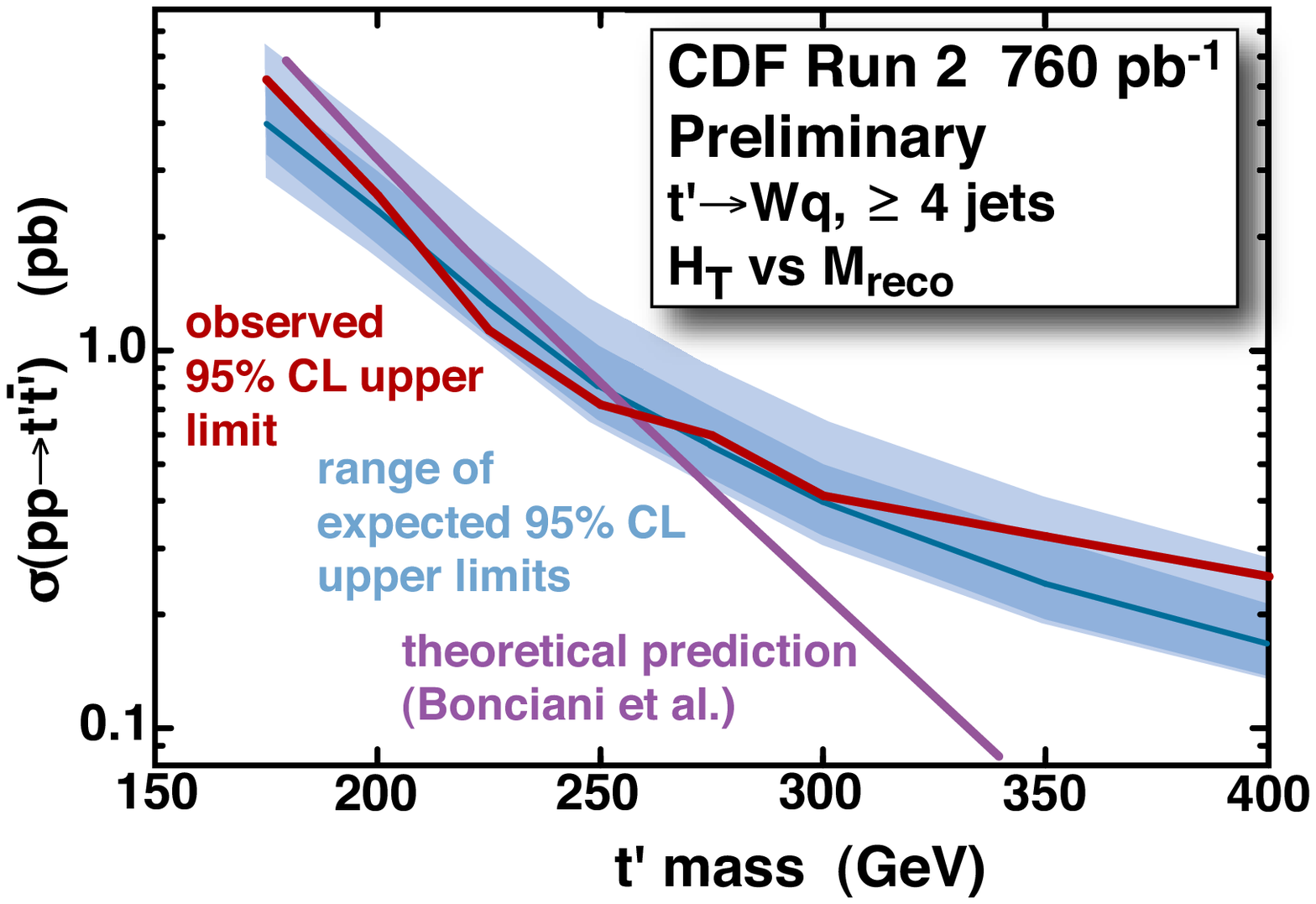}
  \label{restprime}
  \caption{
    {\bf Left:}
    Upper limit, at 95\% C.L., for the production of a heavy narrow resonance decaying into top quark pairs.
    The dots are the observed limits, while the band indicates the $\pm 1$ and $\pm 2$ standard deviation expected limits.
Also shown is a line from a theoretical prediction within a leptophobic topcolor model.
    {\bf Right:}
    Upper limit, at 95\% CL, for the production of a hypothetical fourth generation $t'$ quark as a function of $t'$ quark mass.
    Also shown is the a theoretical calculation for the $t'$ production cross section.
    The band represents $\pm 1$ and $\pm 2$ standard deviation expected limits.
  }
\end{figure}

A narrow ($\Gamma_{X} = 1.2\% M_{X}$) heavy resonance decaying into $t\bar t$ pairs is predicted in various topcolor models like ``topcolor assisted technicolor''\cite{topcolor}.
In some models the resonance couples preferentially to third generation quarks and weakly to leptons.
Searches from Run 1 yield limits in the mass of such a resonance of 480 GeV (CDF) and 560 GeV (D0).
The search is done by looking at the invariant mass $M_{t \bar t}$ distributions of the $t \bar t$ pair.
A resonance would show as a peak in the exponentially falling $M_{t \bar t}$ distribution.
CDF uses a matrix element type of analysis.
In a sample of semi-leptonic top quark candidate events a search is performed for  $M_{t \bar t}>400$ GeV.
No evidence for resonant $t\bar t$ production is observed in 680 pb$^{-1}$ of integrated luminosity.
A small peak previously reported by CDF in the first 320 pb$^{-1}$ has been diluted.
The corresponding preliminary upper cross section limits as a function of $M_{X}$ are shown in Fig.\ \ref{restprime}.
A lower limit of the existence of a $X$ is set to 725 GeV at 95\% C.L.
D0 also performed a search for a narrow resonance in the $t \bar t$ invariant mass spectrum.
A lower limit of the $X$ of 680 GeV at 95\% C.L. is set from an integrated luminosity of 370 pb$^{-1}$.

\subsection{Search for a heavy fourth generation quark}
Several extensions to the SM propose the existence of a heavy fourth generation quark that is not excluded by precision electroweak data or other direct searches \cite{tprime}.
The new quark is referred to here as $t'$, however, it need not to be a standard fourth generation up-type quark.
For the presented analysis it is assumed that this new particle is pair produced via the strong interaction, has a mass greater than the SM top quark, and decays promptly into $Wq$ final states. 
The analysis is performed on a sample of candidate events with a lepton and jets in the final state.
The number of expected background events is compared to the number of events selected in data for distributions of the $t'$ mass, $M_\mathrm{reco}$, and $H_T$.
The background consists of SM $t \bar t$ and $W$+jets.
The mass of the $t'$ quark is reconstructed from a $\chi^2$-fit to the kinematic properties of the final state objects.
The significance of a $t'$ quark signal is extracted from a binned likelihood fit to $H_T$ and  $M_\mathrm{reco}$.
No evidence for the production of $t'$ quark is found.
The upper limits for the production cross section as a function of $t'$ quark mass are shown in Fig.\ \ref{restprime}.
A lower mass limit for $t'$ can be set at $m_{t'}>258$ GeV at 95\% C.L.

\section{Conclusions}

Eleven years after the top quark discovery, we are now in the position to make precision measurements.
The top quark production cross section via the strong interaction has been measured in all major decay channels and is found to be consistent with the SM expectation.
However, the uncertainties on the cross section still allow exotic decay or production mechanisms.
The electroweak production of top quarks (single top) has not been discovered yet.
The current limits are close to the theoretical prediction for the cross section and therefore exclusion or evidence for single top quark production is possible in the near future.
Measurements of the properties of the top quark indicate that its charge is consistent with +2/3$e$ and its lifetime $c\tau$ shorter than $52.5\ \mu$m at 95\% C.L., both consistent with the SM top quark.
One also looks at top quark events to make detailed studies of the electroweak $Wtb$ vertex.
Branching ratios and $W$ boson helicity are found to be consistent with SM expectation.
No evidence for a narrow resonance in top quark decays or additional fourth generation quark is found and limits on the production cross section are set.
Understanding top quark production and properties will be crucial for success at the large hadron collider (LHC).
What we learn at the Tevatron, both in terms of physics and analysis tools, extends directly to the LHC.

%%%%%%%%%%%%%%%%%%%%%%%%%%%%%%%%%%%%%%%%%%%%%%%%
%% BACKMATTER
%%%%%%%%%%%%%%%%%%%%%%%%%%%%%%%%%%%%%%%%%%%%%%%%

\nope{ no ackn, D0 and CDF combined...
\begin{theacknowledgments}
\end{theacknowledgments}
}


\begin{thebibliography}{99}
\bibitem{topdiscovery}
  D0 collaboration, S.~Abachi \etal, Phys.\ Rev.\ Lett. {\bf 74}, 2632 (1995);
  CDF Collaboration, F. Abe \etal, Phys. Rev. Lett. {\bf 74}, 2626 (1995).
\bibitem{massavg} : Tevatron Electroweak Working Group (for the CDF and D0 Collaborations), hep-ex/0608032.
\bibitem{kidonakis} N.~Kidonakis and R. Vogt, Phys. Rev. D {\bf 68}, 114014 (2003),
\bibitem{cacciari} M.~Cacciari \etal, JHEP {\bf 404}, 68 (2004).
\bibitem{cdf} P.~Azzi \etal, Nucl.\ Instrum.\ Method A {\bf 360}, 137 (1995).
\bibitem{d0} V.~Abazov \etal, physics/0507191; Fermilab-Pub-05/341-E.
\bibitem{cdfr} S.~Abachi \etal, Phys.\ Rev.\ Lett {\bf 95}, 102002 (2005).
\bibitem{harris} B.~W.~Harris \etal, Phys.\ Rev.\ D {\bf 66}, 054024 (2002).
\bibitem{poster} see contribution by Shabnam Jabeen, these proceedings.
\bibitem{d0st} V.~Abazov \etal, Phys.\ Lett.\ {\bf 622B}, 265 (2005).
\bibitem{cdfst} S.~Abachi \etal, Phys.\ Rev.\ D {\bf 71}, 012005 (2005).
\bibitem{unki} see contribution by Un-Ki Yang, these proceedings.
\bibitem{chang} D.~Chang \etal, Phys.\ Rev.\ D {\bf 59}, 09153 (1999).
\bibitem{tait} T.~Tait and C.~P.~Yuan, Phys.\ Rev.\ D {\bf 63}, 014018 (2001).
\bibitem{topcolor} C.~T.~Hill, Phys.\ Lett.\ {\bf 266B}, 419 (1991).
\bibitem{tprime} C.~Wagner \etal, hep-ph/0109097; 
  T.~Han \etal, Phys. Lett. {\bf 563B}, 191 (2003);
  H.-J.~He, N.~Polonsky, and S.~Su, hep-ex/0102144;
  L.~Okun \etal, hep-ph/0111028.
\end{thebibliography}
\end{document}